\documentclass[epjCONF]{svjour}
\usepackage{graphics}
\usepackage[varg]{txfonts} 
\usepackage[latin1]{inputenc}

\session-title{A Universe of dwarf galaxies -- Observations, Theories, Simulations}

\begin{document}

%____________________________________________________________

\title{Dwarf galaxies beyond our doorstep: the Centaurus A group}

\author{D. Crnojevi\'{c}\inst{1}\fnmsep\thanks{Corresponding \email{denija@ari.uni-heidelberg.de}. \newline Member of IMPRS (International Max Planck Research School) for Astronomy \& Cosmic Physics at the University of Heidelberg and of the Heidelberg Graduate School for Fundamental Physics.} \and E. K. Grebel\inst{1} \and A. A. Cole\inst{2} \and A. Koch\inst{3} \and M. Rejkuba\inst{4} \and G. Da Costa\inst{5} \and H. Jerjen\inst{5}
}

\institute{Astronomisches Rechen-Institut, Zentrum f\"{u}r Astronomie der
   Universit\"{a}t Heidelberg, M\"{o}nchhofstrasse 12-14, 69120 Heidelberg, Germany 
\and
School of Mathematics \& Physics, University of Tasmania, Private Bag 37 Hobart, 7001 Tasmania, Australia
\and
Department of Physics and Astronomy, University of Leicester, University Road, Leicester LE1 7RH, UK
\and
 European Southern Observatory, Karl-Schwarzschild-Strasse 2, D-85748 Garching, Germany
\and
 Research School of Astronomy \& Astrophysics, Institute of Advanced Studies, 
Australian National University, Cotter Road, Weston Creek, ACT 2611, Australia
}

\abstract{The study of dwarf galaxies in groups is a powerful tool for investigating galaxy evolution, chemical enrichment and environmental effects on these objects. Here we present results obtained for dwarf galaxies in the Centaurus A complex, a dense nearby ($\sim4$ Mpc) group that contains two giant galaxies and about 30 dwarf companions of different morphologies and stellar contents. We use archival optical (HST/ACS) and near-infrared (VLT/ISAAC) data to derive physical properties and evolutionary histories from the resolved stellar populations of these dwarf galaxies. In particular, for early-type dwarfs we are able to construct metallicity distribution functions, find population gradients and quantify the intermediate-age star formation episodes. For late-type dwarfs, we compute recent ($\sim1$ Gyr) star formation histories and study their stellar distribution. We then compare these results with properties of the dwarfs in our Milky Way and in other groups. Our work will ultimately lead to a better understanding of the evolution of dwarf galaxies.
} 

\maketitle

%____________________________________________________________

\section{Introduction} \label{intro}

If the taste of a cake depends primarily on its ingredients, then the study of dwarf galaxies deserves a very careful attention. These objects are indeed believed to be the smallest baryonic counterparts to the dark matter building blocks in our Universe. It is established that dwarf galaxies are by far the most numerous galactic population, and they come in diverse flavours. Early-type dwarfs are predominantly old objects, with a regular, elliptical shape and an extremely low, if not absent, neutral (HI) gas content. The subclasses of this category are dwarf spheroidals, with a lower surface brightness, and dwarf ellipticals, which have higher central concentrations. On the other hand, late-type dwarfs are actively star-forming objects with substantial amounts of neutral gas and irregular shapes (see, e.g., \cite{grebel01}).

The most detailed information we have about dwarf galaxies comes directly from our closest neighbourhood. The Local Group (LG) contains two giant spirals and more than 50 dwarf members, for which decades of excellent data have revealed plenty of properties (e.g., \cite{mateo98,tolstoy09}). Photometry and spectroscopy have been performed on the closest companions of our Milky Way, and perhaps the most intriguing finding from over the last decade has been the extreme diversity in their star formation histories (SFHs). State-of-the-art observations and advanced analysis tools have allowed us to discover a wealth of information, for example that all of the LG dwarfs share a common, ancient population ($\gtrsim10$ Gyr, \cite{grebel04}). The extremes in this sample are, on the one hand, objects that only contain such old populations and, on the other hand, dwarfs that formed most of their stars at more recent times or those that have been experiencing a ``gasping'' star formation over their lives (\cite{marconi95}). LG dwarf galaxies all have a rather low metallicity content ([Fe/H] $\lesssim-1.0$) that is related to their luminosity (e.g., \cite{grebel03}). It is also interesting to analyze the shapes of their metallicity distribution functions (MDFs, e.g., \cite{koch06}), which are fairly well reproduced by theoretical models (e.g., \cite{lanfranchi04,marcolini08}). It has now been shown that many dwarfs, both of elliptical and irregular shape, contain population gradients, with the more metal-rich/younger stars being more centrally concentrated with respect to the more metal-poor/older ones. In some cases the subpopulations are also kinematically distinct (e.g., \cite{harbeck01,tolstoy04,battaglia06}).

In the last few years, many studies have started to look beyond the borders of the LG and have analyzed the properties of dwarf galaxies in nearby groups (e.g., \cite{kara02,trent02,kara04,sharina08,weisz08,bouchard09,dalcanton09}). We are particularly interested in the resolved stellar populations of these objects, since stars provide the record of a galaxy's past evolution and thus constitute a powerful tool to constrain its physical properties. Objects with distances within $\sim10$ Mpc from us are resolvable into individual stars, although the depth of the photometry decreases very rapidly and thus precludes us from reaching the exquisite quality of data we have for LG members. Moreover, as mentioned before, dwarf galaxies are numerous and heterogeneous systems, and in environments different to that of the LG, we can learn more about which factors shape their evolution. In this contribution, we present the first results stemming from the study of the resolved stellar populations of the Centaurus A (CenA) group dwarf members.

The CenA/M83 group is located at a mean distance of $\sim3.7$ Mpc and is composed of two subgroups, each one of which contains a giant galaxy (the peculiar elliptical CenA and the spiral M83) plus a total of about 60 dwarf companions. This group is believed to be a denser and maybe more dynamically evolved system than the LG (see, e.g., \cite{jerjen00a,rejkuba06}), thus making it a very appealing target for dwarf galaxy evolutionary studies and possible environmental effects.

\begin{table*}
 \centering
\caption{Fundamental properties for our sample of early-type galaxies.}
\label{etd}
\begin{tabular}{lccccccccc}
\hline
\hline
Galaxy&RA&DEC&$T$&$D$&$A_{I}$&$M_{B}$&$\Theta$&$<$[Fe/H]$>_{med}$&Last SF\\
&(J2000)&(J2000)&&(Mpc)&&&&(dex)&(Gyr)\\
\hline
{KK189, CenA-dE1}&$13\,12\,45.0$&$-41\,49\,55$&$-3$&$4.42\pm0.33$&$0.22$&$-10.52$&$2.0$&$-1.52\pm0.20$&$\sim9$\\
{ESO269-66, KK190}&$13\,13\,09.2$&$-44\,53\,24$&$-5$&$3.82\pm0.26$&$0.18$&$-13.85$&$1.7$&$-1.21\pm0.33$&$2-3$\\
{KK197, SGC1319.1-4216}&$13\,22\,01.8$&$-42\,32\,08$&$-3$&$3.87\pm0.27$&$0.30$&$-12.76$&$3.0$&$-1.08\pm0.41$&$2-5$\\
{KKs55}&$13\,22\,12.4$&$-42\,43\,51$&$-3$&$3.94\pm0.27$&$0.28$&$-9.91$&$3.1$&$-1.56\pm0.10$&$-$\\
{KKs57}&$13\,41\,38.1$&$-42\,34\,55$&$-3$&$3.93\pm0.28$&$0.18$&$-10.07$&$1.8$&$-1.45\pm0.28$&$-$\\
{CenN}&$13\,48\,09.2$&$-47\,33\,54$&$-3$&$3.77\pm0.26$&$0.27$&$-10.89$&$0.9$&$-1.49\pm0.15$&$-$\\
\hline
\end{tabular}
\begin{list}{}{}
\item[Columns:] (1) name of the galaxy; (2-3) equatorial coordinates (units of right ascension are hours, minutes, and seconds, and units of declination are degrees, arcminutes, and arcseconds); (4) morphological type; (5) distance (derived with the TRGB method); (6) foreground extinction in $I$-band; (7) absolute $B$ magnitude; (8) tidal index (i.e., degree of isolation); (9) median metallicity and metallicity spread; (10) most recent epoch of significant star formation. The references for the reported values are \cite{kara07,crnojevic10,crnojevic10b}.
\end{list}
\end{table*}

%____________________________________________________________

\section{Data} \label{sec:0}

We use archival Hubble Space Telescope (HST) data taken with the Advanced Camera for Surveys (ACS) instrument, which were originally obtained to determine the distance of the targets with the tip of the red giant branch (TRGB) method (see \cite{kara07}). We use data for six early-type dwarfs (companions of CenA, of which four are dwarf spheroidals and two are dwarf ellipticals) and ten late-type dwarfs (five companions of CenA and five of M83). The general properties of our targets are listed in Tab. \ref{etd} and \ref{ltd}. For each of our target dwarfs, observations in the F606W (corresponding to the $V$-band) and F814W (corresponding to the $I$-band) filters are available. These reach approximately 2.5 magnitudes below the TRGB. The photometry on our optical data is run with the DOLPHOT package (\cite{dolphin02}), which we also use to perform extensive artificial star tests in order to quantify the photometric errors and incompleteness for the observations.

For three of our predominantly old early-type dwarf targets, near-infrared (NIR) data have been taken at the Very Large Telescope (VLT) with the Infrared Spectrometer And Array Camera (ISAAC). In combination with the excellent optical data, the images at these wavelengths (specifically, $J$- and $K$-band) help us to distinguish the populations belonging to our targets from the Galactic foreground (the CenA group lies at a galactic latitude of $b\sim20^{\circ}$), and thus to analyze in more detail the stars more luminous than the TRGB, indicative of an intermediate-age population (IAP, with ages in the range $\sim1-9$ Gyr).

%____________________________________________________________

\section{Early-type dwarfs} \label{sec:1}

\subsection{Metallicity distribution functions}

\begin{figure}
 \centering
\resizebox{1\columnwidth}{!}
  {\includegraphics{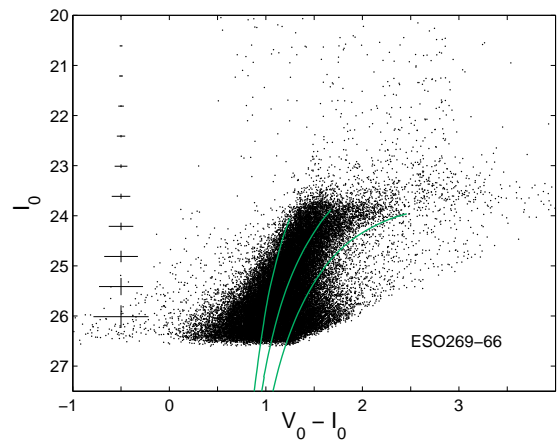}}
 \caption{\footnotesize{Optical (dereddened) CMD for one of the target early-type dwarfs (ESO269-66). The dominant feature of the CMD is a prominent RGB, with a small presence of luminous AGB stars above the TRGB. Representative photometric errorbars are plotted on the left side of the CMD. The green lines are stellar isochrones with a fixed age of 10 Gyr, and metallicity values of [Fe/H] $=-2.5$, $-1.2$ and $-0.4$.}}
 \label{cmd_opt}
\end{figure}

\begin{figure}
 \centering
\resizebox{1\columnwidth}{!}
  {\includegraphics{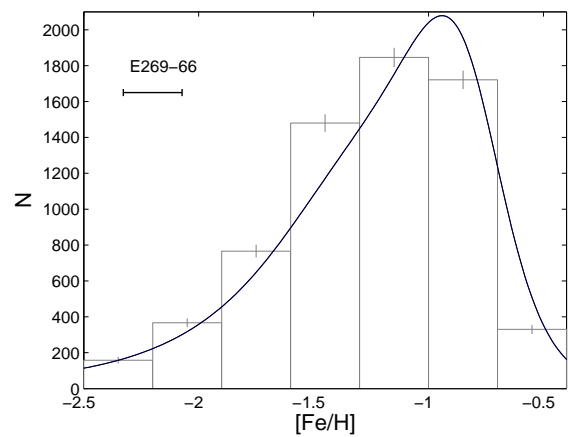}}
 \caption{\footnotesize{MDF for one of the target early-type dwarfs (ESO269-66), derived via interpolation of isochrones at a fixed age (10 Gyr) and with varying metallicity. Overlaid (black line) is the MDF convolved with the observational errors. Also plotted in the upper left corner is the median error on the individual values of [Fe/H].}}
 \label{mdf_opt}
\end{figure}

As can be seen from the example in Fig. \ref{cmd_opt}, the color-magnitude diagrams (CMDs) of our early-type dwarf targets show a prominent red giant branch (RGB), composed of old populations. Although this evolutionary stage suffers from the well known age-metallicity degeneracy, if we only consider a simple stellar population, a spread in metallicity has the effect of producing a broader RGB than a spread in age would do. Given the very little number of luminous AGB stars (indicative of an IAP), we can make the simplified assumption that the stars in our target galaxies are old (we choose a value of 10 Gyr), and assume that the width of their RGBs is entirely due to a range of metallicities (see, e.g., \cite{rejkuba05,harris07}). This is an approximation, but with our data we cannot resolve a possible age spread of a few Gyr at these old ages, which is likely to be present due to prolonged star formation. We use the Dartmouth stellar evolutionary models (\cite{dotter08}), which fit the CMDs of old globular clusters very well (e.g., \cite{glatt08}). We adopt isochrones with varying metallicities, from [Fe/H] $=-2.5$ to $-0.3$ in steps of 0.2 dex, and then interpolate among them to assign a metallicity value to each RGB star. We only do this for stars with magnitudes up to $\sim1.5$ mag fainter than the TRGB, because the photometric errors are smaller and the isochrones are more separated in this region. Moreover, we choose isochrones that have no $\alpha$-element enhancement, since we cannot constrain this. In this way we estimate the mean metallicity for our targets, all of which have a metal-poor stellar content, and draw their MDFs (an example is shown in Fig. \ref{mdf_opt}). We also compute the internal spreads in metallicity for the target dwarfs, after subtraction of the observational errors. Our results are listed in Tab. \ref{etd}.

While a small change ($\sim2$ Gyr) in the chosen age would lead to very small differences in our resulting metallicities, the presence of intermediate-age asymptotic giant branch (AGB) stars could influence the derived MDFs. No population younger than $\sim1$ Gyr is visible in the early-type dwarf CMDs, as can be seen from the absence of blue stars. Although a few luminous AGB stars are definitely present above the TRGB, the relative foreground contamination in this region of the CMD is significant. We give an estimate for the fraction of IAPs in our target objects, and we conclude that the resulting low fractions ($\lesssim20\%$) do not significantly change our results. This and other possible sources of uncertainties are discussed in detail in \cite{crnojevic10}.

We also find weak metallicity gradient as a function of galactic radius in some of our objects. Moreover, we arbitrarily divide the stellar samples into metal-poor and metal-rich (with respect to the median metallicity values), and check whether these subpopulations are similarly distributed within the galaxies. For the two most massive of the objects considered, the metal-rich stars are significantly more centrally concentrated, and a Kolmogorov-Smirnov test confirms that the subsamples are statistically distinct. More than two subpopulations might be present in these galaxies, but this is not resolvable in our observations.

\subsection{Intermediate-age populations} \label{sec:1:iaps}

\begin{figure}
 \centering
\resizebox{1\columnwidth}{!}
  {\includegraphics{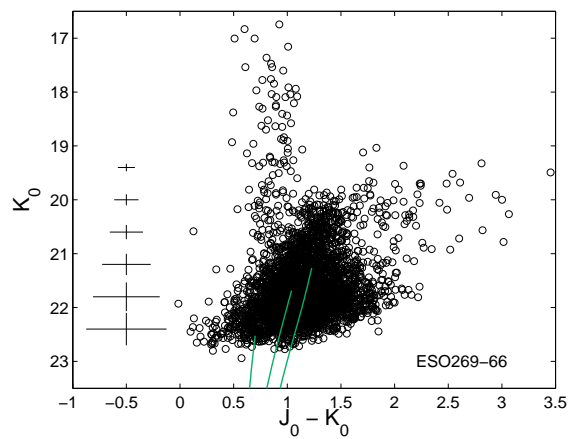}}
 \caption{\footnotesize{NIR (dereddened) CMD for one of the target early-type dwarfs (ESO269-66). As opposed to the optical data, here the Galactic foreground contamination is confined to a vertical sequence (see Section \ref{sec:1:iaps} for details). Representative photometric errorbars are plotted along the CMDs. The green lines are isochrones with a fixed age of 10 Gyr and metallicity values of [Fe/H] $=-2.5$, $-1.2$ and $-0.4$.}}
 \label{cmd_nir}
\end{figure}

We would now like to extract more information about the presence of IAPs in our targets and to estimate the latest epoch when star formation took place there. As mentioned before, it is difficult to constrain the amount of luminous AGB stars in our target galaxies given the amount of Galactic foreground. However, these stars are more luminous at NIR wavelenghts and, most importantly, in a NIR CMD the foreground is clearly confined to a vertical sequence with colors $0.3<J_0-K_0<1.0$ (compare, e.g., Fig. \ref{cmd_nir} to Fig. \ref{cmd_opt}). We thus have a powerful tool, when combining the deep and extremely well resolved optical HST images with the NIR data, to firmly identify candidate luminous AGB stars. We do this by cross-correlating the two stellar catalogs, and by selecting objects found above the TRGB in both datasets (for details, see \cite{crnojevic10b}). We can apply this technique to three of our target early-type dwarfs since we have NIR imaging data for them.
 
From our analysis, we are able to find the most recent episode of star formation by using the empirical relation between the absolute bolometric magnitude of the AGB candidates and their age (see \cite{rejkuba06}). Our results are reported in Tab. \ref{etd}. We compare the derived ratio of luminous AGB stars to RGB stars to the Maraston stellar evolutionary models (\cite{maraston05}), and estimate the IAP fractions for our targets. The resulting fractions are very low, between $5\%$ and $15\%$, thus confirming our previous estimates from the optical data. We note that these fractions are likely to be lower limits to the true ones, given our observational uncertainties and also the uncertainties in the stellar evolutionary models (see, e.g., \cite{melbourne10}).

Finally, we consider the resulting properties found for the whole sample of our target galaxies as a function of deprojected distance from CenA, and as a function of the tidal index (i.e., degree of isolation, positive for galaxies sitting in a dense environment and negative for isolated objects, see e.g. \cite{kara07}). We do not find any obvious correlation between metallicity or IAP fraction and the mentioned parameters, but this result could be biased by the small size of our sample.

%____________________________________________________________

\section{Late-type dwarfs} \label{sec:2}

\subsection{Recent star formation histories}

\begin{table*}
 \centering
\caption{Fundamental properties for our sample of late-type galaxies.}
\label{ltd}
\begin{tabular}{lccccccccc}
\hline
\hline
Galaxy&RA&DEC&$T$&$D$&$A_{I}$&$M_{B}$&$\Theta$&$<$[Fe/H]$>$&$<$SFR$>$\\
&(J2000)&(J2000)&&(Mpc)&&&&(dex)&($10^{-2}$M$_{\odot}$yr$^{-1})$\\
\hline
{ESO381-18}&$12\,44\,42.7$&$-35\,58\,00$&$10$&$5.32\pm0.51$&$0.12$&$-12.91$&$-0.6$&$-1.40\pm0.13$&$0.35\pm0.26$\\
{ESO381-20}&$12\,46\,00.4$&$-33\,50\,17$&$10$&$5.44\pm0.37$&$0.13$&$-14.44$&$-0.3$&$-1.45\pm0.17$&$0.70\pm0.48$\\
{ESO443-09, KK170}&$12\,54\,53.6$&$-28\,20\,27$&$10$&$5.97\pm0.46$&$0.13$&$-11.82$&$-0.9$&$-1.46\pm0.14$&$0.08\pm0.07$\\
{IC4247, ESO444-34}&$13\,26\,44.4$&$-30\,21\,45$&$10$&$4.97\pm0.49$&$0.12$&$-14.07$&$1.5$&$-1.43\pm0.11$&$1.01\pm0.66$\\
{ESO444-78, UGCA365}&$13\,36\,30.8$&$-29\,14\,11$&$10$&$5.25\pm0.43$&$0.10$&$-13.11$&$2.1$&$-1.37\pm0.21$&$0.63\pm0.36$\\
\hline
{KK182, Cen6}&$13\,05\,02.9$&$-40\,04\,58$&$10$&$5.78\pm0.42$&$0.20$&$-12.48$&$-0.5$&$-1.46\pm0.21$&$0.10\pm0.14$\\
{ESO269-58}&$13\,10\,32.9$&$-46\,59\,27$&$10$&$3.80\pm0.29$&$0.21$&$-14.60$&$1.9$&$-0.98\pm0.20$&$6.79\pm3.93$\\
{KK196, AM1318-444}&$13\,21\,47.1$&$-45\,03\,48$&$10$&$3.98\pm0.29$&$0.16$&$-11.90$&$2.2$&$-1.43\pm0.25$&$0.20\pm0.11$\\
{HIPASS J1348-37}&$13\,48\,33.9$&$-37\,58\,03$&$10$&$5.75\pm0.66$&$0.15$&$-11.90$&$-1.2$&$-1.50\pm0.07$&$0.14\pm0.05$\\
{ESO384-16}&$13\,57\,01.6$&$-35\,20\,02$&$10$&$4.53\pm0.31$&$0.14$&$-13.17$&$-0.3$&$-0.97\pm0.15$&$0.60\pm0.23$\\
\hline
\end{tabular}
\begin{list}{}{}
\item[Columns:] (1) name of the galaxy (upper sample: M83 companions, lower sample: CenA companions); (2-3) equatorial coordinates (units of right ascension are hours, minutes, and seconds, and units of declination are degrees, arcminutes, and arcseconds); (4) morphological type; (5) distance (derived with the TRGB method); (6) foreground extinction in $I$-band; (7) absolute $B$ magnitude; (8) tidal index (i.e., degree of isolation); (9) lifetime average SFR; (10) mean metallicity. The references for the reported values are \cite{kara07}, \cite{bouchard09}, \cite{crnojevic10c}, \cite{crnojevic10d}.
\end{list}
\end{table*}

\begin{figure}
 \centering
\resizebox{1\columnwidth}{!}
  {\includegraphics{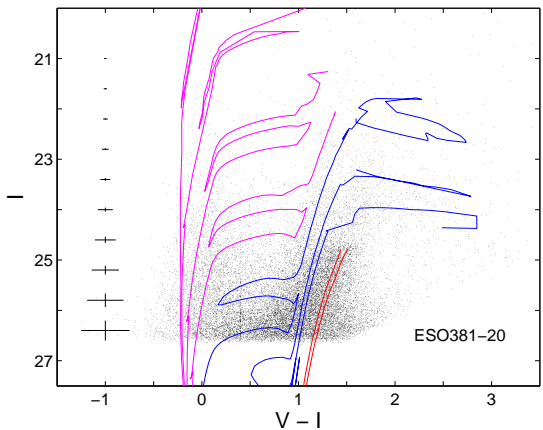}}
 \caption{\footnotesize{Optical CMD for one of the target late-type dwarfs (ESO381-20). We overlay Padova isochrones with a fixed metallicity of [Fe/H] $=-1.4$ and varying ages. The ages are respectively: 4, 12, 35 and 80 Myr (magenta lines, from the blue to the red part of the CMD, indicating massive main sequence and helium-burning stars); 200, 550 Myr, and 1 Gyr (blue lines, helium-burning and AGB stars); 7 and 14 Gyr (red lines, RGB stars). We also report typical $1\sigma$ photometric uncertainties on the left side of the CMD.}}
 \label{cmd_ltd}
\end{figure}

\begin{figure}
 \centering
\resizebox{1\columnwidth}{!}
  {\includegraphics{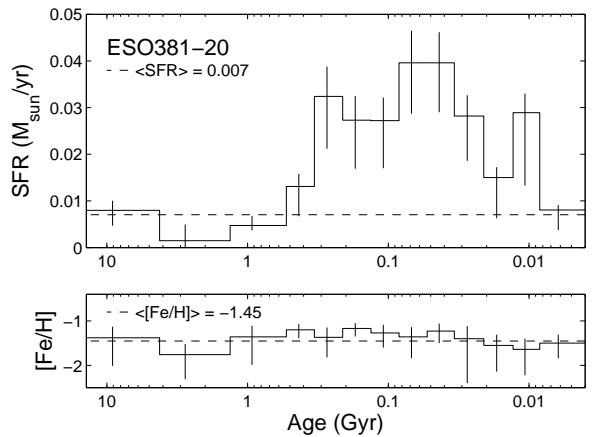}}
 \caption{\footnotesize{\emph{Upper panel}. SFH derived for one of the late-type galaxies studied (ESO381-20). The SFR is plotted as a function of time, the oldest age being on the left side and the most recent time bin on the right edge of the (logarithmic) horizontal axis. The size of the time bins is variable, due to the different amount of information obtainable from each CMD for different stellar evolutionary stages. The black dashed line indicates the mean SFR over the whole galaxy's lifetime. \emph{Lower panel}. Metallicity as a function of time (with the same axis convention as above). The black dashed line represents the mean metallicity over the galaxy's lifetime (note that the metallicity evolution is poorly constrained).}}
 \label{sfh}
\end{figure}

Unlike the early-type dwarfs considered in the previous Section, the late-type dwarfs of our sample show additional structure in their CMDs (see Fig. \ref{cmd_ltd} as an example), thus giving us the possibility to extract more information about their past histories. We divide our targets in two subsamples, namely the M83 and the CenA companions. As can be seen from Tab. \ref{ltd}, some of the dwarfs considered have a positive tidal index, while others have a negative tidal index. We are thus also able to look for environmental effects on their evolution.

We analyze our late-type dwarfs with the synthetic CMD modeling technique, which has proved to be an extremenly powerful tool in deriving the SFHs of dwarf galaxies in the LG and nearby groups (see, e.g., \cite{skillman03,dalcanton09}). Starting from Padova stellar isochrones (\cite{marigo08}, which also reproduce helium-burning stages, as opposed to the Dartmouth ones), we construct synthetic CMDs with a range of physical parameters and then compare them to the observed ones using a maximum likelihood algorithm in order to find the best-fitting solution (for more details, see \cite{cole07,crnojevic10c}). In this way we quantify the amount of stars produced by the galaxy in each different evolutionary stage, which corresponds to a specific age. Given that the only information for ages older than $\sim4-8$ Gyr is coming from the age-metallicity degenerate RGB stage, we are not able to resolve bursts in the star formation for these ages. However, the young, massive main sequence and helium-burning stars, and the intermediate-age luminous AGB stars allow us to reconstruct the star formation rate (SFR) in the past $1-2$ Gyr more accurately. With this technique, it is also possible to derive a mean metallicity estimate for the studied objects, but due to the shallow photometry and the degeneracies present in the CMD, its evolution with time is extremely uncertain. The results of our SFH recovery process are listed in Tab. \ref{ltd} for both subsamples of the galaxies, and show the typical behavior expected for late-type dwarfs.
 
We show an example of a resulting SFH (i.e., star formation rate, SFR, as a function of time) in Fig. \ref{sfh}. The derived SFHs for our sample show considerable diversity in their main features, with some of the target dwarfs having already formed most of their stellar content more than 8 Gyr ago, some others having been significantly active only at recent times, some being possibly in a transition phase from late-type to early-type morphology, and some having just experienced a strong starburst (for more details, see \cite{crnojevic10c,crnojevic10d}). Overall, the general trend seen in all the SFHs is the presence of long quiescent periods, interrupted by more or less extended (few tens to several hundreds Myr) episodes of enhanced star formation.

\subsection{Stellar spatial distributions}

As a final step, we divide the stellar content of each galaxy into subsamples with different ages and produce density maps for each of them. It is interesting to see a confirmation of previous studies (e.g., \cite{dohm97,glatt08,weisz08}), that the youngest stars tend to be concentrated in clumps close to the central regions of the galaxies, while the oldest stars are uniformely distributed all over their bodies. This stems from the fact that the young stars still have the imprint position of their birthplaces, while the old ones have had the time to migrate and redistribute within the galaxy.

We take advantage of the intrinsic properties of the blue helium-burning (BHeB, forming a vertical sequence with $V-I\sim0$, see Fig. \ref{cmd_ltd}) stars to estimate the star formation timescales. In this evolutionary stage, stars with different ages are very well separated in the CMD, as opposed to what is seen, for example, in the RGB phase. We can thus trace the spatially resolved SFH of a galaxy by considering stellar maps of BHeB stars with different ages. We find that star forming complexes have diameters of $\sim100$ pc and timescales of $\gtrsim100$ Myr, and we confirm that global bursts in a galaxy are long-lived (several hundreds of Myr) events, within which smaller, localized star forming regions form and dissolve (see also \cite{mcquinn09}).

Our results point towards a stochastic star formation mode for late-type dwarfs (see also \cite{weisz08}). We also look for possible relations between the physical properties of our dwarfs and their environment. There is no clear correlation between the lifetime average SFR with tidal index, or with deprojected distance from the closest giant galaxy, and the same conclusion is also valid for the metallicity. This is a reasonable result, if we consider that average properties depend on the whole galaxy's history, and we only have information about the current position of the objects within the group because their orbits are unknown. However, if we compute the ratio of the SFR in the last $\sim500$ Myr to the lifetime average value, we do find that it correlates with environment. Namely, dwarfs that are closest to the dominant group galaxy, and found in a denser environment, have a lower such ratio with respect to the more isolated objects (see also \cite{bouchard09}). Finally, we find that dwarfs located in denser regions will consume their HI gas content much faster than isolated ones, provided they continue to form stars at their lifetime average SFR.

%____________________________________________________________

\section{Discussion and conclusions} \label{concl}

We have studied a sample of dwarf galaxies in the nearby CenA/M83 group, an environment similar to the LG but denser (in terms of number of galaxies) and possibly more evolved. The results reported in this contribution can be now compared to what we know about our own LG.

In terms of metallicity content of their old populations, the early-type dwarfs studied are metal-poor and always show metallicity spreads, just as the LG companions. They moreover follow the luminosity-metallicity relation originally seen for the LG (see, e.g., \cite{grebel03}), and then also found to extend to other nearby environments (e.g., \cite{sharina08}). The shapes of the derived MDFs are overall very similar to the ones spectroscopically derived for LG dwarf members (e.g., \cite{koch06}), which implies the combination of supernovae-driven enrichment and galactic outflows, although we note that a more detailed comparison is not viable because of the intrinsic uncertainties of our method. Some of our target dwarfs also show the presence of distinct stellar populations, similarly to some LG dwarfs (e.g., Fornax, Sculptor, Sextans). The intriguing difference between the two groups emerges only when we look at the IAPs more carefully. Namely, an analysis of the luminous AGB stars tells us that the IAP fractions in CenA companions are smaller than what is found for Milky Way dwarf spheroidal companions and for M31 dwarf elliptical companions (the relative differences are not affected by model uncertainties). Our results are more similar to what is seen for dwarf spheroidal companions of M31 (see \cite{crnojevic10b}). This is one of the first studies where IAPs for dwarfs in nearby groups are investigated from their resolved stellar populations. Although our sample is still too small to draw firm conclusions, our results suggest that the CenA environment could somehow have played a role in the suppression of its companions' star formation. This is a very interesting result which certainly deserves more observational evidence.

The late-type dwarfs considered in our study all show complex and varied SFHs, with lifetime average SFRs of between $\sim10^{-3}$ and $\sim6 \times 10^{-2}$M$_\odot$yr$^{-1}$. If we compare these values to LG dwarf galaxies in the same luminosity range, we find that the CenA/M83 dwarfs have slightly higher values, comparable to what is found for the M81 group (a highly interacting nearby group) late-type dwarfs (see \cite{weisz08}). The global enhancements in SFR that we are able to resolve from the recent SFHs of our targets last for several hundreds of Myr, producing stars at a rate of $\sim2-3$ times the average SFR. We show that within these bursts, localized star forming regions have sizes of $\sim100$ pc and timescales of $\sim100$ Myr. We conclude that star formation in these objects is a stochastic process. Finally, also among late-type dwarfs we find a hint for environmental effects, in the sense that dwarfs in denser environments and closer to the dominant galaxy have had lower SFRs in the last $\sim500$ Myr, and are consuming their gas reservoirs faster than the isolated ones.

The CenA/M83 group has proved to be a very promising target, in which the resolved stellar populations of the dwarf companions can be analyzed to better understand the details of galactic evolution. The study presented in this contribution should thus be an appetizer, and stimulate the curiosity for the work that has started to be done for nearby groups of galaxies.

%____________________________________________________________

\vspace{-0.3cm}

\begin{acknowledgement}

DC wishes to thank the organizers of the conference for making this meeting stimulating and enjoyable. DC acknowledges travel support from the IMPRS and ARI/ZAH. DC is grateful to S. Pasetto and S. Jin for a careful reading of the manuscript and for their support.

\end{acknowledgement}

\vspace{-0.6cm}

%____________________________________________________________

%____________________________________________________________

\end{document}